\documentclass{article}

\usepackage{arxiv}
\usepackage[utf8]{inputenc} 
\usepackage[T1]{fontenc}    
\usepackage{hyperref}       
\usepackage{url}            
\usepackage{booktabs}       
\usepackage{amsfonts}       
\usepackage{nicefrac}       
\usepackage{microtype}      
\usepackage{lipsum}		
\usepackage{graphicx}
\usepackage{natbib}
\usepackage{doi}
\usepackage{amsmath}
\usepackage{listings}
\usepackage{pythonhighlight}

\usepackage[ruled,vlined]{algorithm2e}

\SetKwInput{KwInput}{Input}
\SetKwInput{KwOutput}{Output}
\SetKw{In}{in}
\SetKw{And}{and}
\SetKw{Mod}{mod}
\SetKw{Exists}{exists}

\title{Steganography of Complex Networks}

\author{ 
\href{https://orcid.org/0000-0002-3004-2901}{\includegraphics[scale=0.06]{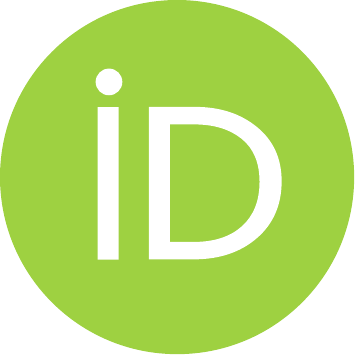}\hspace{1mm}
Daewon~Lee} \\
School of Art and Technology\\
Chung-Ang University\\
Anseong, South Korea\\
\texttt{dwlee@cau.ac.kr} 
}

\hypersetup{
pdftitle={Steganography of Complex Networks},
pdfauthor={Daewon~Lee},
pdfkeywords={Information Hiding, Steganography, Complex Network, Network Topology},
}

\begin{document}
\maketitle

\begin{abstract}
Steganography is one of the information hiding techniques, which conceals secret messages in cover media. Digital image and audio are the most studied cover media for steganography. However, so far, there is no research on steganography to utilize complex networks as cover media. To investigate the possibility and feasibility of complex networks as cover media for steganography, we introduce steganography of complex networks through three algorithms: BIND, BYMOND, and BYNIS. BIND hides two bits of a secret message in an edge, while BYMOND encodes a byte in an edge, without changing the original network structures. Encoding simulation experiments for the networks of Open Graph Benchmark demonstrated BIND and BYMOND can successfully hide random messages in the edge lists. BYNIS synthesizes edges by generating node identifiers from a given message. The degree distribution of stego network synthesized by BYNIS was mostly close to a power-law. Steganography of complex networks is expected to have  applications such as watermarking to protect proprietary datasets, or sensitive information hiding for privacy preservation.
\end{abstract}

\begin{figure}[ht]
\centering
\includegraphics[width=0.5\linewidth]{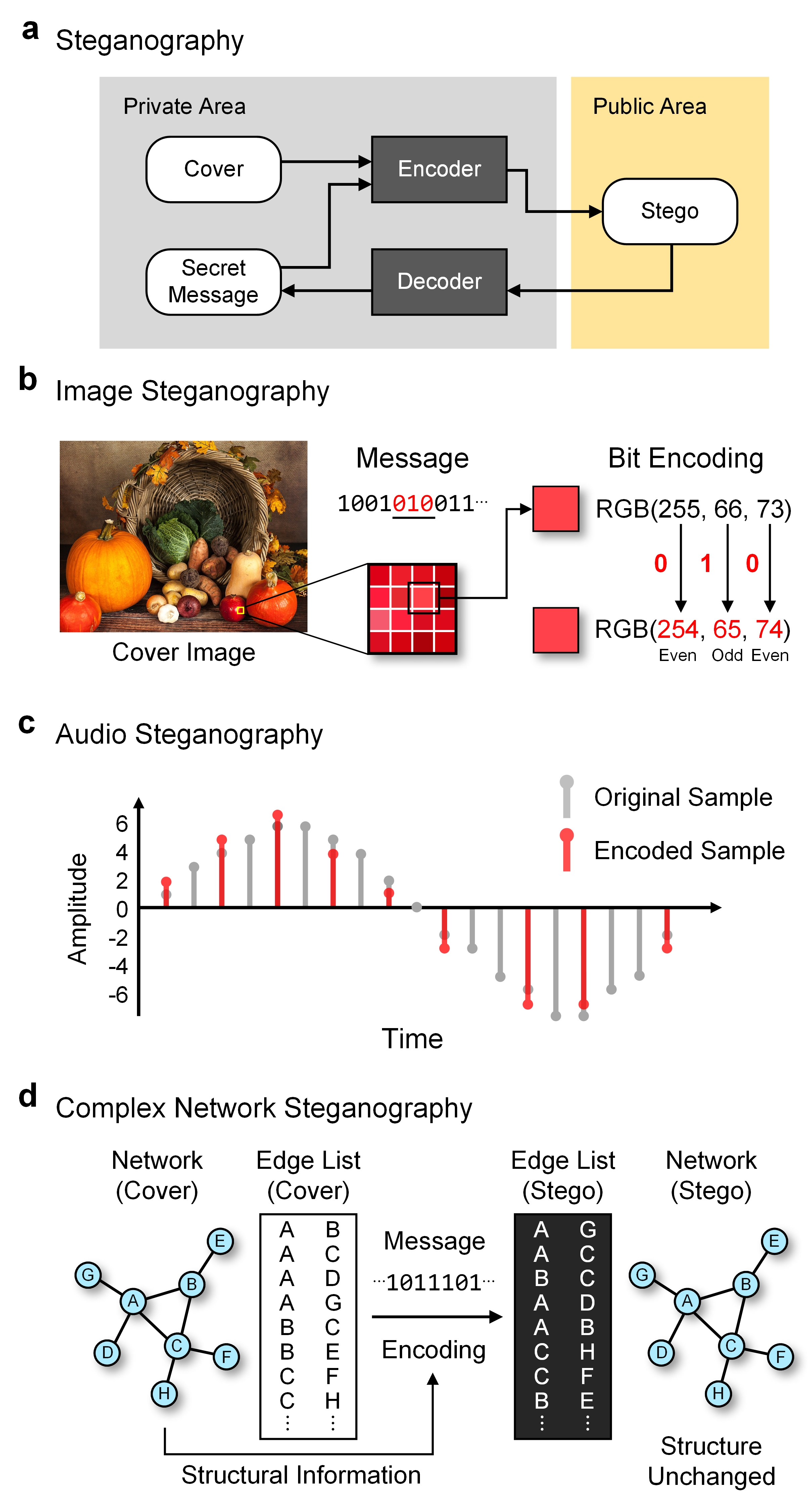}
\caption{\textbf{Illustration of steganography and applications to various media.}}
\label{fig:intro_steganography}
\end{figure}

Steganography is a type of invisible communications that hide secret messages in media \cite{Fridrich2009book, Subhedar2014}. One of the important goals in steganography is to hide the existence of secret message as well as the message itself \citep{Cheddad2010}. Steganography has been seriously studied since terrorists, spies, and hackers have been suspected of utilizing steganography to conceal secret messages for their malicious purposes until recently \citep{Homer-Dixon2002, Hosmer2006, Zielinska2014}. In modern digital steganography, a steganographic algorithm basically involves encoder and decoder (Fig. \ref{fig:intro_steganography}a). The encoder hides message bits in a medium, called \textit{cover}. The encoded medium that contains the message, called \textit{stego}, can be conveyed through public communication channels, as it is usually difficult to distinguish stegos from covers by human perception. The role of decoder is to recover the secret message from the stego without loss of information. Image and audio files are the most popular covers for steganography. For instance, Least-significant-bit (LSB) embedding for image steganography is the simplest algorithm, which embeds message bits in the LSB of each RGB color pixel in a lossless image format (Fig. \ref{fig:intro_steganography}b). The modified pixel is interpreted as a bit according to the parity of pixel value in the decoding process. This LSB embedding method can also be applied to audio steganography, where encoder hides the message bits in audio samples, as in RGB pixels (Fig. \ref{fig:intro_steganography}c).

Scientific data is also gaining attention as scalable cover media for steganography today, since the size of scientific data varies from kilobytes to petabytes.
A representative example is DNA steganography.
Yachie \emph{et~al.} and Shipman \emph{et~al.} successfully demonstrated the genomes of living organisms can encode digital data, which allows us to adopt a variety of genomes to store and hide information in DNA bases ranging from millions of bacterial bases to billions of human bases \citep{yachie2007, shipman2017}.
Clelland \emph{et~al.} first introduced a DNA steganography, in which DNA triplet represents a single alphabetic letter \citep{clelland1999}.
Na devised a method that hides secret messages in the variable regions of genome (single nucleotide polymorphisms) to evade detection  \citep{na2020}.
Li \emph{et~al.} developed an experimental method using CRISPR/Cas12a system, which takes advantage of the specific and non-specific primers in polymerase chain reaction (PCR) as real and fake keys, respectively, to enhance security \citep{li2018}.
Moreover, in recent years, a DNA steganalysis based on deep learning has been developed to detect DNA steganography \citep{bae2020}.

In addition to DNA-based steganography, various methods have shown steganography of scientific data.
Kim \emph{et~al.} introduced a steganography based on an immuno-chemical system, where immuno-specific interactions on the ELISA plate result in the combinations of colors that encode text messages \citep{kim2011}.
Sakar \emph{et~al.} developed a chemical-based method that hides secret messages within the emission spectra of a unimolecular fluorescent sensor \citep{sarkar2016message}.
Boukis \emph{et~al.} demonstrated a secret chemical communication system, where the molecular keys are decoded by high resolution tandem mass spectrometry, and the decoded messages can be used as passwords for digital encryption algorithms such as AES \citep{boukis2018multicomponent}.
Purcell \emph{et~al.} developed an experimental method that hides the original topology of a synthetic
gene circuit by camouflaging the circuit, and recovers the original circuit by adding molecular keys to remove the activity of genes in the camouflaged circuit \citep{purcell2018}.
Zhang \emph{et~al.} experimentally demonstrated a protein binding-based steganography based on DNA origami cryptography \citep{zhang2019}.

However, so far, there is no research on steganography using network structure data as cover media.
A variety of phenomena in the real world can be represented and analyzed in complex networks, and the field of network science has emerged to systematically analyze networks \citep{barabasi2014book}.
As graph theory and network science advance, some satisfactory solutions for complex network problems such as traffic system\citep{dijkstra1959}, search engine\citep{page1999}, terrorism\citep{carley2002}, etc. have been obtained.
Recently, deep neural networks for graphs, called graph neural networks (GNN), are bringing the power of artificial intelligence to network science \citep{wu2020}.
Recognizing the importance of systematic evaluation, a benchmarking dataset including carefully curated networks of different sizes, types, and tasks has been opened for developing and comparing the performance of machine learning models \citep{hu2020}.
Despite the advent of various types of real-world network datasets and corresponding analytical models, steganography for complex networks has not been studied.

To investigate the possibility and feasibility of complex networks as cover media for information hiding, we introduce a novel steganography for network datasets (Fig. \ref{fig:intro_steganography}d). We developed three steganographic algorithms: \textit{BIND}, \textit{BYMOND}, and \textit{BYNIS}. BIND and BYMOND algorithms hide secret messages in existing network data files such as edge lists, without changing the original topology of the networks. BIND encodes the bits of a given message into the node degrees of an edge in an edge list, while BYMOND encodes the message bytes, rather than the bits, into the node degrees. We demonstrate BIND and BYMOND algorithms can successfully hide randomly generated messages in the real-world network datasets of Open Graph Benchmark (OGB) \citep{hu2020}.
On the other hand, BYNIS creates a synthetic network for a given message, while the degree distribution of the synthetic network is close to power-law distribution. We expect that steganography of complex networks will have important applications such as watermarking to protect proprietary datasets of a company, or sensitive information hiding for privacy preservation of patients.
 
\section*{Results}

\subsection*{Steganographic Algorithms for Real-World Networks}

One of the important goals of steganographic algorithms is to hide a secret message in the existing data.
Therefore, we developed two steganographic algorithms for real-world network datasets: BIND and BYMOND (Fig. \ref{fig:bind} and Fig. \ref{fig:bymond}).
The format of input and output in both algorithms is an edge list, which is the most common and simplest data format for real-world networks.

\subsubsection*{BIND algorithm}

\begin{table}[ht]
\centering
\caption{\label{tab:ogb} \textbf{Open Graph Benchmark datasets for validating steganographic algorithms}.
To validate the steganographic algorithms for complex networks, we selected various sizes of edge lists from Open Graph Benchmark (OGB) datasets.
The names in parentheses are the original ID of OGB.
The items are sorted by the number of edges.
}
\vspace{3mm}
\begin{tabular}{|c|c|c|c|p{50mm}|}

\hline
No. & Dataset & Num. Nodes (|V|) & Num. Edges (|E|) & Description  \\
\hline
1 & 
ddi (ogbl-ddi) &
4,267 &
1,334,889 &
A undirected network of drug-drug interactions \cite{wishart2018drugbank}. \\
\hline
2 &
arxiv (ogbn-arxiv) &
169,343 &
1,166,243 &
A directed network of citations between all computer science papers of arXiv indexed by Microsoft Academic Graph (MAG) \cite{wang2020microsoft}. \\
\hline
3 & collab (ogbl-collab) & 235,868 & 1,285,465 &
A undirected network of collaborations between authors indexed by MAG \cite{wang2020microsoft}.
\\
\hline
4 & wikikg2 (ogbl-wikikg2) & 2,500,604 & 17,137,181 & 
A knowledge graph from Wikidata knowledgebase \cite{vrandevcic2014wikidata}. \\
\hline
5 & ppa (ogbl-ppa) & 576,289 & 30,326,273 & 
A undirected network of protein-protein associations \cite{szklarczyk2019string}. \\
\hline
6 & citation2 (ogbl-citation2) & 2,927,963 & 30,561,187 &
A directed network of citations between papers indexed by MAG \cite{wang2020microsoft}.
\\
\hline
7 & proteins (ogbn-proteins) & 132,534 & 39,561,252 & 
A undirected network of protein-protein associations. \cite{szklarczyk2019string}. \\
\hline
8 & products (ogbn-products) & 2,449,029 & 61,859,140 &
A undirected network of Amazon co-purchase \cite{bhatia2016}. \\
\hline
\end{tabular}

\end{table}

\begin{figure}[ht]
\centering
\includegraphics[width=0.85\linewidth]{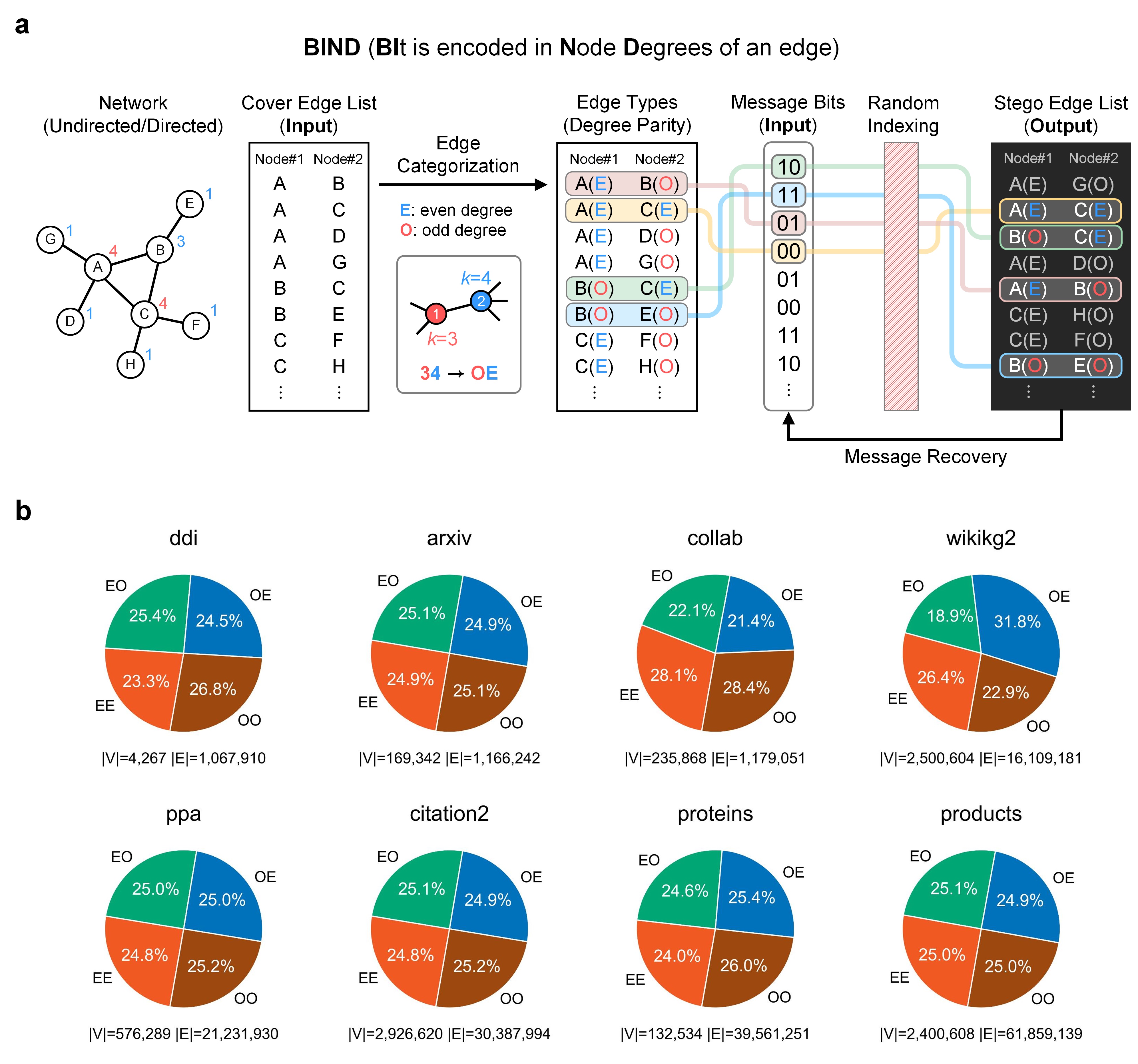}
\caption{\textbf{BIND algorithm.} \textbf{(a)} Schematic diagram of BIND algorithm.
A single bit of secret message is encoded into parity of node degree.
\textbf{(b)} Proportions of the four edge types in OGB datasets. $|V|$ and $|E|$ represent the number of nodes and the number of edges, respectively, counted in each edge list. 
$|V|$ and $|E|$ can be different from the original values in Table \ref{tab:ogb}, as BIND utilizes the edge list of a raw format.}
\label{fig:bind}
\end{figure}

\textbf{BIND} (\textbf{BI}t is encoded in \textbf{N}ode \textbf{D}egrees of an edge) encodes the bits of a given message into a list of edges according to node parity (Fig. \ref{fig:bind}a).
First, BIND categorizes each edge according to the parities of two node degrees.
For instance, if the first node has degree 3 and the second node has degree 4,
the edge is categorized as “OE”, which represents an edge of the “Odd” and “Even” degrees (Fig. \ref{fig:bind}a, the edge with k=3 and k=4).
The important point in this process is that the order of appearance of the nodes must be distinguished in an edge.
So, "OE" and "EO" are categorized as different edge types.
BIND then reads the message bits in units of two bits, and matches the corresponding edge.
For instance, "01" in the message bits corresponds to an edge between node A and node B, whose type is "EO" (Fig. \ref{fig:bind}a, the edge shown in red shade).
Finally, stego edges are arranged by random indexing with a seed that is usually assigned by user password.
Only the order of edges in the stego edge list is different from the order of edges in the cover edge list.
Recovering the original message from the stego edge list of BIND is reversing the random indexing with the seed, and sequentially interpreting the stego edge types as message bits.

The real-world networks with the evenly distributed edge types are promising covers for BIND algorithm, if we assume that lossless compression and cryptographic algorithms can randomize the message bits following a uniform distribution \citep{bassham2010, klein2020}.
Hence, we analyzed the edge categorization of BIND for OGB datasets (Fig. \ref{fig:bind}b).
Interestingly, the proportions of the four edge types are almost the same in each dataset except "wikikg2".
The distribution of edges in "wikikg2" is slightly biased towards the "OE" type.
These results imply that real-world networks exemplified by OGB datasets are suitable for BIND algorithm, if we assume that the 2-bit patterns in secret messages are uniformly distributed.

\begin{figure}[tbp]
\centering
\includegraphics[width=0.6\linewidth]{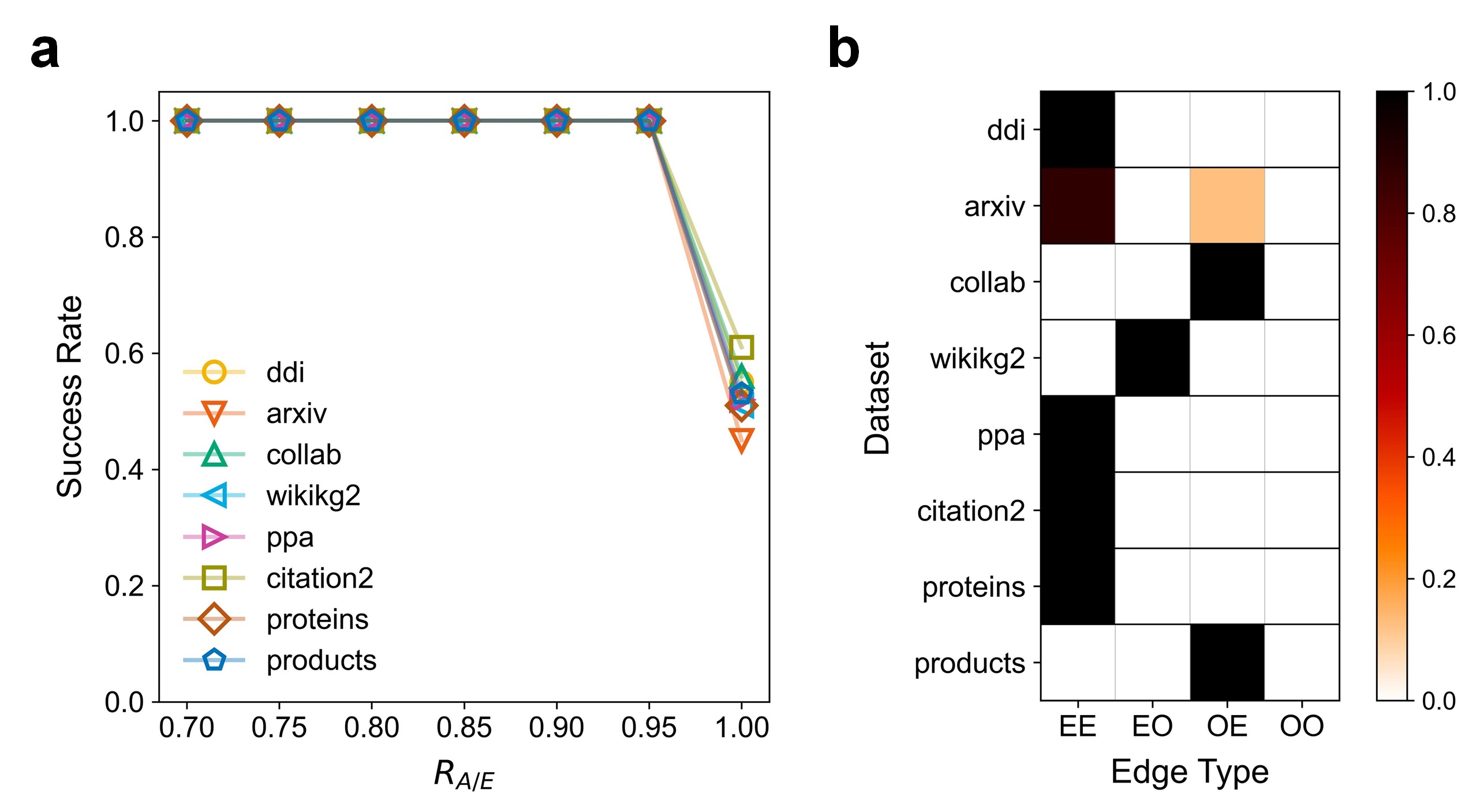}
\caption{\textbf{Encoding simulation experiments for BIND algorithm.}
\textbf{(a)} We randomly generated messages according to $R_{A/E}$, and observed whether the steganographic algorithm had successfully completed encoding the given messages for OGB datasets. $R_{A/E}$ is the ratio of the actual number of message bits to the estimated maximum number of message bits. Success rate represents the fraction of successful completions in the 100 simulation experiments.
\textbf{(b)} To understand the success rate decreases under the condition of $R_{A/E}=1.0$, we also counted the failed cases of each edge type in the simulation experiments. The color represents the ratio of failed cases to the total failed cases for each edge type.
}
\label{fig:bind_inc_payload}
\end{figure}

To analyze the payload capacity of complex network steganography, we defined a payload capacity measure, named BPE (Bits Per Edges), as follows:
\begin{equation}
BPE := \frac{|B_{msg}|}{|E|}\notag\\
\end{equation}
where $|B_{msg}|$ is the number of message bits, and $|E|$ is the number of edges in the edge list.
As a single edge encodes two bits in BIND, the maximum BPE of BIND is theoretically calculated as follows:
\begin{equation}
BPE_{max}^{thr} := \frac{|B_{max}^{thr}|}{|E|} = \frac{2\cdot|E|}{|E|} = 2\notag\\
\end{equation}
where the theoretical maximum number of the message bits, $|B_{max}^{thr}|$, is equal to $2\cdot|E|$, which means all edges encode the message bits in the edge list.
However, the minimum size of edge set among the sets of the four edge types determines the lower bound of the maximum BPE in BIND, since the encoding is impossible if the edges of a certain edge type corresponding to a 2-bit pattern are insufficient. Therefore, the lower bound of $BPE_{max}$ can be described as follows:
\begin{equation} 
|B_{max}| \ge 4\cdot|E_{min}|
\quad
\Rightarrow \quad 
BPE_{max} = \frac{|B_{max}|}{|E|} \ge \frac{4\cdot|E_{min}|}{|E|}\notag\\
\end{equation}
where $E_{min}$ is the set of the minimum size among the sets of the four edge types in BIND and $|E_{min}|$ is the size of $E_{min}$. We can use the lower bound to estimate actual $|B_{max}|$ and $BPE_{max}$ of a given edge list if we do not know the exact $BPE_{max}$ as follows:
\begin{equation}
|B_{max}^{est}| := 4\cdot|E_{min}|
\quad \Rightarrow \quad 
BPE_{max}^{est} := \frac{4\cdot|E_{min}|}{|E|}\notag\\
\end{equation}
where $|B_{max}^{est}|$ and $BPE_{max}^{est}$ are the estimates of $|B_{max}|$ and $BPE_{max}$, respectively.
The $BPE_{max}^{est}$ can be utilized as a basic measure for evaluating and comparing the payload capacities of complex network covers.
We also defined $R_{A/E}$ to efficiently analyze and control the message size as follows:
\begin{align}
R_{A/E} := \frac{|B_{msg}|}{|B_{max}^{est}|} \quad
\Rightarrow \quad |B_{msg}| = R_{A/E}\cdot|B_{max}^{est}|
= R_{A/E}\cdot(4\cdot|E_{min}|)
\label{eq:bind_msg_size}
\end{align}
where the subscript, $A/E$, represents a ratio between the actual number of message bits and the estimated maximum.

We performed simulation experiments to analyze the payload capacity in BIND (Fig. \ref{fig:bind_inc_payload}).
We repeated the encoding simulation 100 times for random messages and measured the success rate.
Based on equation (\ref{eq:bind_msg_size}), random messages of $|B_{msg}|$ for simulation were generated following a uniform distribution.
BIND successfully encoded random messages, but the success rate dramatically decreased for $R_{A/E}$ = 1.0 (Fig. \ref{fig:bind_inc_payload}a).
In the process of encoding the messages generated with $R_{A/E}$ = 1.0, the number of edges of a particular type corresponding to a 2-bit pattern could be insufficient with a high probability, since the message size,
$|B_{msg}| = |B_{max}^{est}|$, was very close to the actual maximum size.
To understand that the success rate decreases under the condition of $R_{A/E}$ = 1.0, we also counted the failed cases of each edge type in the encoding simulation experiments for BIND algorithm.
We found that almost all failed cases are attributed to the lack of $E_{min}$ edges (Table \ref{tab:bind_ogb} and Fig. \ref{fig:bind_inc_payload}b, dark cells).
These results imply that the payload size must be carefully determined in BIND algorithm, and we can use $R_{A/E}$ less than 1.0 to control the payload capacity.

\begin{table}[tbp]
\centering
\caption{\label{tab:bind_ogb}
\textbf{Estimation of payload capacity in BIND}.
{\boldmath${E_{min}}$}: the set of minimum size among the sets of the 4 edge types in BIND.
{\boldmath$|E_{min}|$}: the size of $E_{min}$;
{\boldmath$|B_{max}^{est}|$}: the estimated maximum number of the message bits,
which is calculated by $4 \cdot |E_{min}|$;
{\boldmath$|B_{max}^{thr}|$}: the theoretical maximum number of the message bits;
{\boldmath$R_{E/T}$}: the ratio of $|B_{max}^{est}|$ to $|B_{max}^{thr}|$ (i.e., $|B_{max}^{est}|/|B_{max}^{thr}|$), which indicates how close the estimate is to the theoretical value.
}
\vspace{3mm}

\begin{tabular}{|c|c|c|c|c|c|}
\hline
Dataset &
Type of $E_{min}$  &
$|E_{min}|$ &
$|B_{max}^{est}|$ &
$|B_{max}^{thr}|$ &
$R_{E/T}$ \\
\hline
ddi & EE & 248,568 & 994,272 & 1,067,910 & 0.931 \\
\hline
arxiv & EE & 290,238 & 1,160,952 & 1,166,242 & 0.995 \\
\hline
collab & OE & 252,044 & 1,008,176 & 1,179,051 & 0.855 \\
\hline
wikikg2 & EO & 3,043,263 & 12,173,052 & 16,109,181 & 0.756 \\
\hline
ppa & EE & 5,266,145 & 21,064,580 & 21,231,930 & 0.992 \\
\hline
citation2 & EE & 7,541,565 & 30,166,260 & 30,387,994 & 0.993 \\
\hline
proteins & EE & 9,500,440 & 38,001,760 & 39,561,251 & 0.961 \\
\hline
products & OE & 15,426,359 & 61,705,436 & 61,859,139 & 0.998 \\
\hline
\end{tabular}

\end{table}

\subsubsection*{BYMOND algorithm}

\begin{figure}[htb]
\centering
\includegraphics[width=0.85\linewidth]{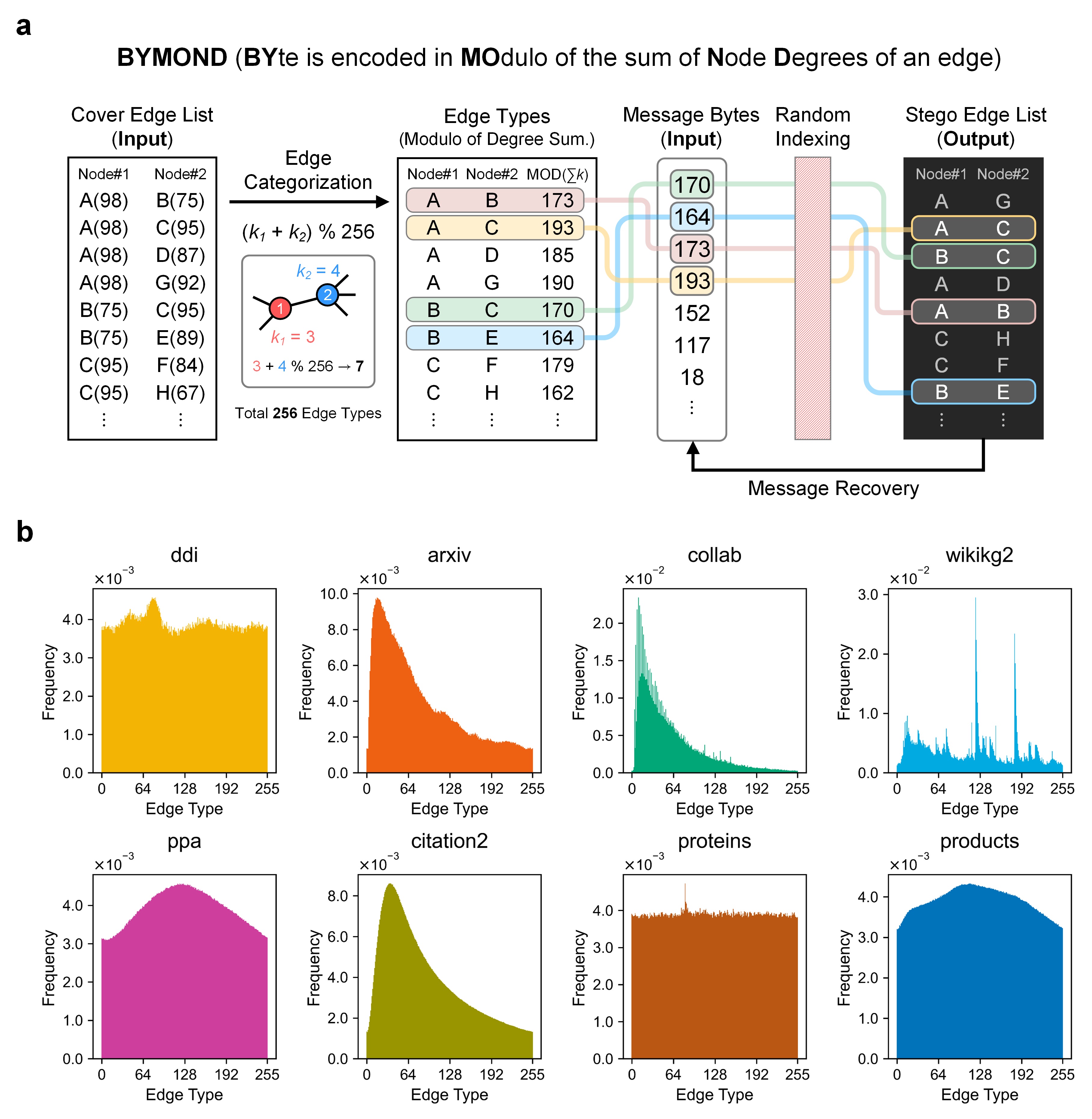}
\caption{\textbf{BYMOND algorithm.} \textbf{(a)} Schematic diagram of BYMOND algorithm.
A single byte of secret message is encoded into the modulo of the sum of node degrees.
\textbf{(b)} Frequency of the 256 edge types in OGB datasets. $|V|$ and $|E|$ represent the number of nodes and the number of edges, respectively.}
\label{fig:bymond}
\end{figure}

\begin{figure}[tbp]
\centering
\includegraphics[width=1.0\linewidth]{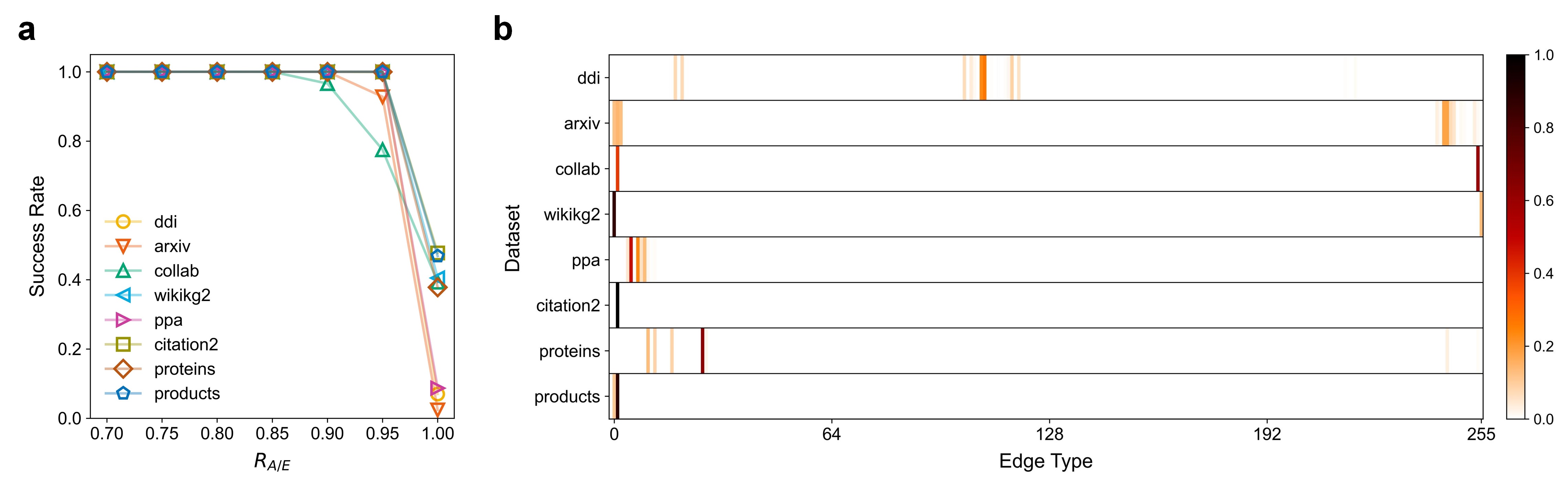}
\caption{\textbf{Encoding simulation experiments for BYMOND algorithm.} 
All the same as in Figure \ref{fig:bind_inc_payload}, except that the number of edge types in BYMOND is 256, and simulation experiment was repeated 1,000 times
(see the main text for details).
}
\label{fig:bymond_inc_payload}
\end{figure}

To improve the payload capacity, we developed \textbf{BYMOND}(\textbf{BY}te is encoded in \textbf{MO}dulo of the sum of \textbf{N}ode \textbf{D}egrees of an edge) algorithm (Fig. \ref{fig:bymond}a).
Instead of encoding two bits, an edge encodes a byte (8 bits) in BYMOND. As one byte can have 256 different values (i.e., 0 to 255), BYMOND categorizes edges into 256 types according to the modulo of the sum of node degrees.
For instance, if an edge with $k=3$ and $k=4$ is categorized as 7, then the edge encodes a single byte, 7 (Fig. \ref{fig:bymond}a).
The reason for applying modulo operation to the sum of degrees is that the sum of degrees can exceed 255, which is the maximum value of a byte.
The rest of encoding and decoding processes in BYMOND is the same as BIND algorithm, except that an edge encodes a byte.
Figure \ref{fig:bymond}b shows the results of edge categorization in BYMOND.
Unlike the results of BIND, the edge types of BYMOND are not evenly distributed, but have a variety of distributions.
For example, the distributions of "arxiv", "collab", and "citation2" are skewed towards the edge types of lower values.
However, the edges of "protein" dataset are almost evenly distributed compared to the other datasets.

We also performed the encoding simulation experiments for BYMOND, where the simulation was repeated 1,000 times for random messages.
In BYMOND, $B_{max}^{est}$ and $|B_{msg}|$ are determined as follows:
\begin{align}
|B_{max}^{est}| = 256\cdot|E_{min}|
\quad \Rightarrow \quad 
|B_{msg}| = R_{A/E}\cdot(256\cdot|E_{min}|)
\label{eq:bymond_msg_size}
\end{align}
where 256 is the number of edge types in BYMOND.
The encoding success rates of BIND rate ranged from 0.4 to 0.6 for OGB datasets when $R_{A/E}=1.0$ (Fig. \ref{fig:bind_inc_payload}a),
whereas the success rates of BYMOND were different depending on the dataset (Fig. \ref{fig:bymond_inc_payload}a).
BYMOND failed to encode random messages under $R_{A/E}=1.0$  for "ddi", "arxiv" and "ppa" with a high probability.
For the other datasets such as "collab", "wikikg2", "citation2", "proteins" and "products", the success rates of BYMOND were between 0.35 and 0.5.
Figure \ref{fig:bymond_inc_payload}b shows the ratio of failed cases to the total failed cases for each edge type in the 1,000 simulations.
The encoding failures for "collab", "wikikg2", "citation2", "proteins" and "products" were mainly due to the lack of $E_{min}$ edges 
(Table \ref{tab:bymond_ogb} and Fig. \ref{fig:bymond_inc_payload}b, dark bars).
However, the failures for "ddi", "arxiv" and "ppa" datasets were attributed to the lack of several edge types (Fig. \ref{fig:bymond_inc_payload}b, several peach and orange bars).
These results suggest that the probability of encoding failure increases in BYMOND when the sizes of multiple edge sets in a network dataset are close to $|E_{min}|$.
Therefore, the distribution of edge types and $|E_{min}|$ should be carefully considered to select cover networks for BYMOND algorithm.

\begin{table}[tbp]
\centering
\caption{\label{tab:bymond_ogb}
\textbf{Estimation of payload capacity in BYMOND}.
{\boldmath$E_{min}$}: the set of minimum size among the sets of the 256 edge types in BYMOND;
{\boldmath$|B_{max}^{est}|$}: the estimated maximum number of the message bits,
which is calculated by $256 \cdot |E_{min}|$;
the others are the same as those in Table \ref{tab:bind_ogb}.}
\vspace{3mm}
\begin{tabular}{|c|c|c|c|c|c|}
\hline
Dataset &
Type of $E_{min}$&
$|E_{min}|$ &
$|B_{max}^{est}|$ &
$|B_{max}^{thr}|$ &
$R_{E/T}$  \\
\hline
ddi & 109 & 3,804 & 7,790,592 & 8,543,280 & 0.912 \\
\hline
arxiv & 253 & 1,536 & 3,145,728 & 9,329,936 & 0.337 \\
\hline
collab & 254 & 275 & 563,200 & 9,432,408 & 0.060 \\
\hline
wikikg2 & 0 & 20,206 & 41,381,888 & 128,873,448 & 0.321 \\
\hline
ppa & 9 & 65,625 & 134,400,000  & 169,855,440 & 0.791 \\
\hline
citation2 & 1 & 39,703 & 81,311,744 & 243,103,952 & 0.334 \\
\hline
proteins & 26 & 149,477 & 306,128,896 & 316,490,008 & 0.967 \\
\hline
products & 1 & 197,578 & 404,639,744 & 494,873,112 & 0.818 \\
\hline
\end{tabular}
\end{table}

\subsection*{Steganography by Network Synthesis}

\subsubsection*{BYNIS algorithm}

\begin{figure}[tbp]
\centering
\includegraphics[width=0.78\linewidth]{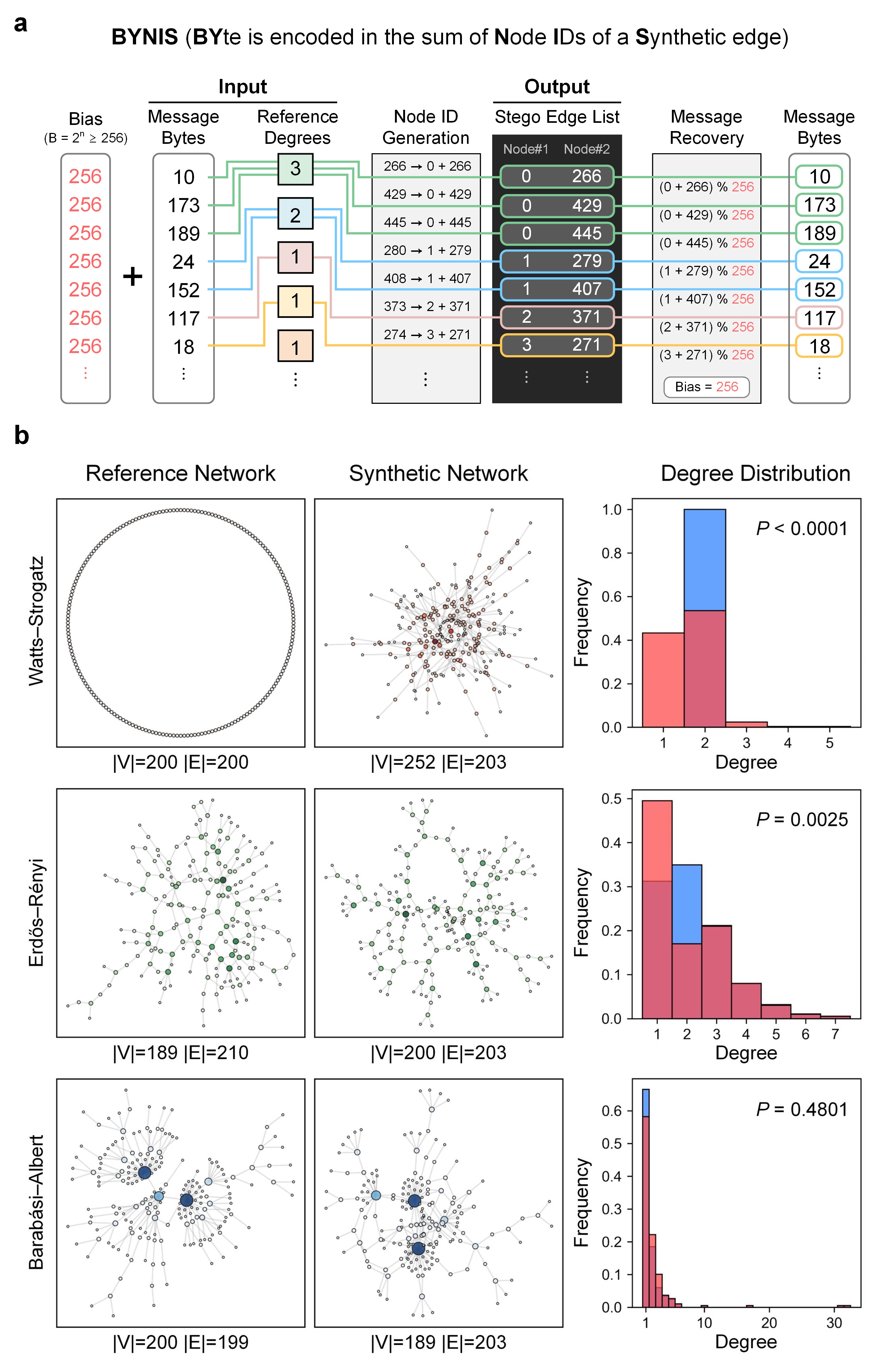}
\caption{\textbf{BYNIS algorithm.} \textbf{(a)} Schematic diagram of BYNIS algorithm.
An edge is synthesized by splitting a single byte of secret message into two node identifiers of the edge.
\textbf{(b)} Structures and degree distributions of synthetic networks created by BYNIS referring to random network models. \textit{P} represents \textit{p}-value of Kolmogorov-Smirnov test for two samples, where null hypothesis is that two samples are drawn from the same distribution.}
\label{fig:bynis}
\end{figure}

BIND and BYMOND algorithms utilize existing real-world network datasets.
However, high payload capacity is not guaranteed if the degree distribution of cover network does not conform to the bit or byte patterns of message data (Fig. \ref{fig:bind_inc_payload} and Fig. \ref{fig:bymond_inc_payload}).
Hence, we developed a steganographic algorithm, named BYNIS(\textbf{BY}te is encoded in the sum of \textbf{N}ode \textbf{I}Ds of a \textbf{S}ynthetic edge), which synthesizes the edges of complex networks according to a given message (Fig. \ref{fig:bynis}).
First, BYNIS splits a byte into two integers, which represent the two node identifiers (IDs) of an edge.
In contrast to BYMOND, BYNIS algorithm encodes a message byte into the sum of node IDs, not node degrees.
A bias can be added to the message bytes to prevent the failure of generating node IDs when the byte values of message are too small.
BYNIS can also refer to reference degrees for generating node IDs.
The node ID generation algorithm is a kind of greedy algorithm,
which generates a node ID whose degree is currently the maximum.
This greedy algorithm sequentially divides message bytes to generate a specific node ID until the reference degree for the specific node ID is exhausted to create synthetic edges.
For instance, if the first degree is 3 in reference degrees,
BYNIS splits the first 3 bytes into 3 pairs of node IDs that must have node ID "0" (Fig. \ref{fig:bynis}a). 
As the node degree 3 is exhausted for node ID "0", the next node ID is "1".
To recover the original message from the stego edge list of BYNIS,
we should apply the modulo operation to the sum of the node IDs as follows:
\begin{equation}
B_{i}^{rec} := (\text{id}_{1} + \text{id}_{2})\ mod\ \text{bias} \notag\\
\end{equation}
where $B_{i}^{rec}$ is the $i$-th recovered byte, and id\textsubscript{1} and id\textsubscript{2} represent the two node IDs of the $i$-th synthetic edge in a stego edge list.
In Figure \ref{fig:bynis}, for instance, the 7-th edge is decoded as follows:
\begin{equation}
B_{7}^{rec} = (3 + 271)\ mod\ 256\ = 18 \notag\\
\end{equation}

To characterize synthetic networks created by BYNIS,
we investigated how synthetic networks are created according to the reference degrees of different random network models.
Given three reference degrees generated from Watts-Strogatz, Erdős-Rényi, and Barabási-Albert models,
the network synthesized with Barabási-Albert model was most similar to the original random network (Fig. \ref{fig:bynis}b, p-value > 0.1 for two degree distributions).
In other words, splitting message bytes into node IDs based on the greedy approach for edge generation in BYNIS creates a network, whose degree distribution is close to a power law distribution
\citep{barabasi1999emergence}.

\section*{Discussion}

We introduce steganography of complex networks by demonstrating possibility and feasibility of three steganographic algorithms through analysis of real-world network datasets.
BIND algorithm encodes two bits of a given secret message into two degrees of an edge according to node degree parity, while BYMOND algorithm encodes a single byte of message into the sum of node degrees of an edge adjusted by modulo operation.
BYNIS algorithm synthesizes edges by splitting message bytes into node identifiers based on a greedy approach, resulting in a synthetic network whose degree distribution follows a power-law.

As payload capacity is one of the most important aspects in steganography,
we define $BPE$ (Bits Per Edges), and explain how to estimate $BPE_{max}$ with $|B_{max}^{est}|$ and $|E_{min}|$. In the ideal case, the number of a specific bit or byte pattern in secret message exactly matches the number of edges for each edge type of BIND or BYMOND.
However, it is usually hard to find a real-world network that perfectly conform to a secret message data.
In general, $|E_{min}|$ limits the maximum payload capacity, $|B_{max}^{est}| = \textit{\text{the number of edge types}}\cdot|E_{min}|$ can guide the payload capacity of a given cover network for BIND or BYMOND algorithm.

Encoding simulation experiments for OGB datasets show that BIND and BYMOND can fail to encode messages of high capacity that are close to the maximum capacity of message, $|B_{max}|$.
Since the distribution of the edge types of BYMOND in each OGB dataset do not follow uniform distribution (Fig. \ref{fig:bymond}b), the encoding success rate of BYMOND varied depending on dataset in the simulation experiments (Fig. \ref{fig:bymond_inc_payload}a).
In other words, BYMOND is more sensitive to edge type distribution than BIND, although the payload capacity of BYMOND is theoretically 4 times larger than that of BIND.
To make the distribution of edge types follow the uniform distribution, we can define a new function that determines edge types based on node degrees as follows:
\begin{equation}
B_{i} := f(k_{1}, k_{2})  \notag\\
\end{equation}
where $B_{i}$ is the \textit{i}-th message byte, and $k_{1}$ and $k_{2}$ are two node degrees of an edge.
In BYMOND, $f$ is defined as follows:
\begin{equation}
f(k_{1}, k_{2}) := (k_{1}+k_{2})\ mod\ 256  \notag\\
\end{equation}
A well-designed function $f$ is expected to achieve both evenly distributed edge types and high payload capacities for a variety of real-world networks, overcoming the limitations of BYMOND.
Steganography of synthetic networks is exemplified by BYNIS, which is essentially a hybrid algorithm that integrates text steganography\citep{taleby2019modern} with the structural information of complex network. In other words, the node identifier characters of an edge are generated to create a synthetic network, referring to a degree distribution.
To reflect the properties of real-world networks to synthetic networks, BYNIS utilizes the degree distribution of a reference network.
However, BYNIS does not exactly reflect any degree distribution as shown in the experiments of random network models, except power-law distribution (Fig. \ref{fig:bynis}b).
Hence, networks with power-law distributions should be preferred to other distributions in order to take advantage of the BYNIS algorithm.
Recently, Broido \textit{et al.} has demonstrated that scale-free networks are empirically rare, and log-normal distributions (i.e., $\frac{1}{x}e^{-\frac{(log{x}-\mu)^2}{2\sigma^2}}$) are more appropriate than power laws to explain degree distributions of most real-world networks \citep{broido2019}.
Therefore, the edge synthesis algorithm, currently based on a greedy approach in BYNIS, can be improved to reproduce various degree distributions including log-normal as well as power-law.
We expect steganography of complex networks to have useful applications. As networks in industry such as social networks or product networks are becoming important assets \citep{haenlein2011, carmi2017}, steganographic algorithms for complex networks can be adopted as watermarking techniques to protect the proprietary datasets.
Another important application is to hide sensitive information in network datasets. For instance, when we need to publicly distribute patient-specific cancer networks \citep{drake2016}, we can hide personal information of a patient in his or her network data without changing the original structure of the network.
We hope that this study will be the first step to facilitate the development of various algorithms and applications based on steganography of complex networks.

\newpage

\section*{Methods}

\subsection*{Algorithm implementation}

All steganographic algorithms have been implemented in Python programming language.
To enhance performance of algorithms, we use NumPy\citep{harris2020array} and Pandas\citep{pandas2010} packages, which enable us to utilize high performance vectorization in array programming.
We can choose between NetworkX\citep{networkx2008} and igraph\citep{igraph2006} to handle the data structures of undirected and directed networks in our implementation.
bitstring\citep{griffiths2020} is adopted to efficiently process bits and bytes of message data.
We provide the core part of each algorithm written in pseudo-code as follows:

\begin{algorithm}[!htb]
  \KwInput{cover network $G_{c}$; cover edge list $E_{c}$; message bits $M_{b}$; password $P$}
  \KwOutput{stego edge list $E_{s}$}

  \tcp{Edge categorization}
  \ForEach{ $edge$ \In $E_{c}$ }
  {
     $node_{1}$, $node_{2}$ $\gets$ $edge$ \\
     $D_{1}$ $\gets$ Degree($G_{c}$, $node_{1}$) \\
     $D_{2}$ $\gets$ Degree($G_{c}$, $node_{2}$) \\
     
     \If{$(D_{1}\ \Mod\ 2 = 0)$ \And $(D_{2}\ \Mod\ 2 = 0)$}{
       Append($E_{ee}$, $edge$) \\
     }
     \ElseIf{$(D_{1}\ \Mod\ 2 = 0)$ \And $(D_{2}\ \Mod\ 2 = 1)$}{
       Append($E_{eo}$, $edge$) \\
     }
     \ElseIf{$(D_{1}\ \Mod\ 2 = 1)$ \And $(D_{2}\ \Mod\ 2 = 0)$}{
       Append($E_{oe}$, $edge$) \\
     }
     \ElseIf{$(D_{1}\ \Mod\ 2 = 1)$ \And $(D_{2}\ \Mod\ 2 = 1)$}{
       Append($E_{oo}$, $edge$) \\
     }
  }
   
  \tcp{Message encoding}
  \ForEach{ \textup{two bits} $b_{1}, b_{2}$ \In $M_{b}$ }
  {
     \If{$(b_{1}\ \Mod\ 2 = 0)$ \And $(b_{2}\ \Mod\ 2 = 0)$}{
      $edge$ $\gets$ Pop($E_{ee})$ \\
    }
    \ElseIf{$(b_{1}\ \Mod\ 2 = 0)$ \And $(b_{2}\ \Mod\ 2 = 1)$}{
      $edge$ $\gets$ Pop($E_{eo})$ \\
    }
    \ElseIf{$(b_{1}\ \Mod\ 2 = 1)$ \And $(b_{2}\ \Mod\ 2 = 0)$}{
      $edge$ $\gets$ Pop($E_{oe})$ \\
    }
    \ElseIf{$(b_{1}\ \Mod\ 2 = 1)$ \And $(b_{2}\ \Mod\ 2 = 1)$}{
      $edge$ $\gets$ Pop($E_{oo})$ \\
    }
    Append($E_{s}$, $edge$) \\
   }
   
   \tcp{Randomization of edge sequence}
   Seed($P$) \\
   Randomize($E_{s}$) \\
   \KwRet $E_{s}$

  \caption{Message encoding algorithm of BIND}
\end{algorithm}

\begin{algorithm}[!htb]
  \KwInput{cover network $G_{c}$; cover edge list $E_{c}$; message bytes $M_{B}$; password $P$}
  \KwOutput{stego edge list: $E_{s}$}
   \tcp{Edge categorization}
   \ForEach{ $edge$ \In $E_{c}$ }
   {
      $node_{1}$, $node_{2}$ $\gets$ $edge$ \\
      $D_{1}$ $\gets$ Degree($G_{c}$, $node_{1}$) \\
      $D_{2}$ $\gets$ Degree($G_{c}$, $node_{2}$) \\
      $t$ $\gets$ $(D_{1} + D_{2})$ \Mod 256 \\
      Append($E[{t}]$, $edge$) \\
   }
   
   \tcp{Message encoding}
   \ForEach{ \textup{one byte} $B$ \In $M_{B}$ }
   {
      $t$ $\gets$ $B$
      edge $\gets$ Pop($E[{t}]$, $edge$) \\
      Append($E_{s}$, $edge$) \\
   }
   
   \tcp{Randomization of edge sequence}
   Seed($P$) \\
   Randomize($E_{s}$) \\
   
  \KwRet $E_{s}$

  \caption{Message encoding algorithm of BYMOND}
\end{algorithm}

\newpage

\begin{algorithm}[!htb]
  \KwInput{reference degrees $D_{r}$; message bytes $M_{B}$}
  \KwOutput{stego edge list $E_{s}$}
  \tcp{$D_{r}$ is assumed to be a sorted array}
   $bias \gets 256$ \\
   $M_{B}$ $\gets$ $M_{B} + bias$ \quad\tcp{Add a bias to message bytes.}
   $D$ $\gets$ ZerosLike($D_{r}$)\quad\tcp{Zero-initialized array like $D_{r}$}
   $i$ $\gets$ 0 \quad\tcp{Current node ID}
   \ForEach{ \textup{one byte} $B$ \In $M_{B}$ }
   {   
      \If{$D[i]$ < $D_{r}[i]$}{
        \tcp{Consume the current node degree.}
        $D[i]$ $\gets$ $D[i] + 1$
      } \Else {
        \tcp{Update to the next node.}
        $i$ $\gets$ $i + 1$ \\
      }
      \tcp{Greedy approach: select a node whose degree is currently the maximum.}
      $id_{1}$ $\gets$ $i$ \quad\tcp{$i$ represents the node that has the max degree.}
      $id_{2}$ $\gets$ $B - i$ \\
      
      \tcp{Create a new edge with node IDs.}
      $edge$ $\gets$ ($id_{1}$, $id_{2}$) \\

      \tcp{Change the second node ID if the synthetic edge already exists.}
      $j$ $\gets$ 1 \\
      \While{$edge$ \Exists \In $E_{s}$} {
        $id_{2}$ $\gets$ $id_{2} + bias*j$ \\
        $edge$ $\gets$ ($id_{1}$, $id_{2}$) \\
        $j$ $\gets$ $j + 1$ \\
      }
      
      Append($E_{s}$, $edge$) \\
   }
  \KwRet $E_{s}$

  \caption{Network synthesis algorithm of BYNIS}
\end{algorithm}

\newpage

\subsection*{Network datasets}

Open Graph Benchmark (OGB) provides real-world networks, which we use to validate our steganographic algorithms.
We can download the OGB datasets using OGB python package.
When creating a dataset object such as \texttt{PygNodePropPredDataset} or \texttt{PygLinkPropPredDataset}, OGB package downloads each dataset if it does not exists in a local storage.
The following Python source code is an example for downloading the datasets.

\begin{python}
from ogb.nodeproppred import PygNodePropPredDataset
from ogb.linkproppred import PygLinkPropPredDataset

list_nodeproppred_datasets = [
    "ogbn-arxiv",
    "ogbn-proteins",
    "ogbn-products"    
]

list_linkproppred_datasets = [
    "ogbl-ddi",
    "ogbl-collab",
    "ogbl-wikikg2",
    "ogbl-ppa",
    "ogbl-citation2",
]

dpath_download = "/data/ogb/"

for name in list_nodeproppred_datasets:
    dataset = PygNodePropPredDataset(name=name, root=dpath_download)

for name in list_linkproppred_datasets:
    dataset = PygLinkPropPredDataset(name=name, root=dpath_download)
\end{python}

In the directory of each OGB dataset, we only use the edge list of a raw format (e.g., \path{/data/ogb/ogbl_ddi/raw/edge.csv}).
Some datasets including \texttt{ogbl-collab} have redundant rows as multiple edges in their raw edge lists.
All redundant edges in the edge list are maintained and utilized for message encoding. 
However, BIND and BYMOND assume all edges are unique in a directed graph when interpreting the properties of network structure.
So, the number of edges in a edge list and the number of edges in a network structure can be different.

\subsection*{Encoding simulation experiments}

We performed encoding simulation experiments to understand the encoding algorithms of BIND and BYMOND for random messages.
In the experiments for BIND,
we increased $R_{A/E}$ from 0.7 to 1.0 by 0.5,
and performed 100 simulations for each $R_{A/E}$.
In a single simulation, message size, $|B_{msg}|$, was determined by equation (\ref{eq:bind_msg_size}).
In the experiments for BYMOND, 
the simulation conditions were the same as in BIND except the number of simulations and the message size.
To catch encoding failures as many as possible,
we set the number of simulations greater than the number of edge types, 256.
The number of simulations was set $1,000$ ($\ge 256$),
and the message size was determined by equation (\ref{eq:bymond_msg_size}) in BYMOND.

\subsection*{Network synthesis with random network models}
To generate reference degrees, we used random network generation functions of NetworkX\cite{networkx2008}.
The following example shows the functions and parameters we used for generating the reference networks in Figure \ref{fig:bynis}.
We computed Kolmogorov-Smirnov static to compare the degree distributions between reference network and stego network.

\begin{python}
import networkx as nx
import scipy as sp

list_ref = []
num_nodes = 200

# Watts-Strogatz
g = nx.newman_watts_strogatz_graph(num_nodes, 2, 0.01)
list_ref.append(g)

# Erdos-Renyi
g = nx.fast_gnp_random_graph(int(1.2*num_nodes, 0.008, directed=False)
list_ref.append(g)

# Barabasi-Albert
g = nx.barabasi_albert_graph(num_nodes, 1)
list_ref.append(g)

# Get the giant component
for i, g in enumerate(list_ref):
    nodes_gc = sorted(nx.connected_components(g), key=len, reverse=True)
    g_ref = g.subgraph(nodes_gc[0]) 
    
    # ... omitted for brevity
    
    # BYNIS algorithm
    g_stego = encode(msg_bytes, g_ref)
    
    degrees_ref = get_degrees(g_ref)
    degrees_stego = get_degrees(g_stego)
    
    # Compare two samples using Kolmogorov-Smirnov test
    res = sp.stats.ks_2samp(degrees_ref, degrees_stego)
    print("P = 
\end{python}

\section*{Data availability}
To obtain the network datasets in this study, we need to download OGB datasets (\url{https://ogb.stanford.edu}). Refer to "Network datasets" in "Methods" section for details.

\section*{Code availability}
We provide a GitHub repository for the steganographic algorithms: \url{https://github.com/dwgoon/sgcn}

\newpage

\bibliographystyle{unsrtnat}
\bibliography{references}  

\end{document}